\documentclass[10pt,a4paper,twocolumn]{article} 

\usepackage[utf8]{inputenc}
\usepackage{lmodern} % science advances submission 
\usepackage[scale=0.97]{sourcesanspro}
\usepackage{times} % science advances submission 
\usepackage{geometry}
\usepackage[backend=biber, 
  style=science,
  isbn=false,
  articletitle=true,
  maxnames=10,
  ]{biblatex}
\usepackage{csquotes} % avoid warning
\AtEveryBibitem{\clearfield{day}}
\AtEveryBibitem{\clearfield{month}}
\AtEveryBibitem{\clearlist{language}}

\DeclareBibliographyDriver{software}{%
  \usebibmacro{bibindex}%
  \usebibmacro{begentry}%
  \usebibmacro{author/translator+others}%
  \setunit{\printdelim{nametitledelim}}\newblock
  \printfield{title}\ \printfield{version}%
  \newunit
  \usebibmacro{date}%
  \newunit\newblock
  \printlist{organization}%
  \newunit
  \usebibmacro{publisher+location+date}%
  \newunit\newblock
  \printfield{doi}%
  \newblock
  \ (\printfield{url})%
  \newunit\newblock
  \iftoggle{bbx:related}
    {\usebibmacro{related:init}%
     \usebibmacro{related}}
    {}%
  \usebibmacro{finentry}}

\usepackage{amsmath,amssymb}
\usepackage{array} % improved formatting and options for equation arrays
\DeclareMathAlphabet{\mathup}{OT1}{\familydefault}{m}{n} % only for non-lualatex

\newcommand*{\mup}[1]{\mathup{#1}} % Abkürzung für \mathup
\newcommand*{\diff}{\,\mathup{d}}
\newcommand*{\iu}{\mathup{i}\mkern1mu}
\DeclareRobustCommand{\Re}{\operatorname{Re}}

\DeclareMathOperator{\e}{e}
\newcommand\ten[1]{\mathbf{#1}}
\newcommand\dirvec[1]{\mathbf{e}_{#1}}

\usepackage[ngerman,USenglish]{babel}
\makeatletter
\adddialect\l@en\l@USenglish
\adddialect\l@Englisch\l@USenglish
\adddialect\l@de\l@ngerman
\makeatother

\usepackage{siunitx}
\DeclareSIUnit{\permille}{\text{\textperthousand}}

\usepackage{graphicx}
\DeclareGraphicsExtensions{.pdf,.jpeg,.png}
\usepackage[hidelinks]{hyperref}

\title{Beating resonance patterns and extreme power flux skewing in anisotropic elastic plates} 
\author
{Daniel~A.~Kiefer,$^{\ast}$ Sylvain~Mezil, Claire~Prada\\
\\
\normalsize{Institut Langevin, ESPCI Paris, Université PSL, CNRS, 75005 Paris, France}\\
\normalsize{$^\ast$To whom correspondence should be addressed; E-mail:  daniel.kiefer@espci.fr.}
}
\date{}

\begin{document}

\makeatletter
\twocolumn[
\begin{@twocolumnfalse} 
\vspace{-0.4cm}
\maketitle 
\centering
\begin{abstract}\noindent\normalfont\sffamily
  Elastic waves in anisotropic media can exhibit a power flux that is not collinear with the wave vector. This has notable consequences for waves guided in a plate. Through laser-ultrasonic experiments, we evidence remarkable phenomena due to slow waves in a single crystal silicon wafer. Waves exhibiting power flux orthogonal to their wave vector are identified. A pulsed line source that excites these waves reveals a wave packet radiated parallel to the line. Furthermore, there exist precisely eight plane waves with zero power flux. These so-called zero-group-velocity modes are oriented along the crystal's principal axes. Time acts as a filter in the wave vector domain that selects these modes. Thus, a point source leads to beating resonance patterns with moving nodal curves on the surface of the infinite plate. We observe this pattern as it emerges naturally after a pulsed excitation.\\\medskip\noindent
  Published version: Science Advances 9 (51), p. eadk6846, doi \href{https://dx.doi.org/10.1126/sciadv.adk6846}{10.1126/sciadv.adk6846}.

\end{abstract}\bigskip
\end{@twocolumnfalse}
]
\makeatother

\section{Introduction}
Engineering materials often exhibit anisotropic stiffness, in particular, single crystals and composites. Among them, monocrystalline silicon is the single most important material for fabrication of integrated circuits, microelectromechanical systems (MEMS) and photovoltaic cells~\cite{hopcroft_what_2010,belyaev_crack_2006}. These are generally produced from a thin wafer of this material. Understanding the intricate mechanics of elastic wave propagation in these plates is not only of importance for their quality evaluation~\cite{chakrapani_crack_2012,song_crack_2013,masserey_defect_2019}, but also for the functional design of MEMS devices such as surface acoustic wave (SAW) and Lamb wave filters and sensors~\cite{hackett_aluminum_2023,yantchev_thin-film_2011,caliendo_zero-group-velocity_2017}. The latter usually involve layers of piezoelectric materials, which constitute another kind of medium where anisotropy plays a major role~\cite{rupitsch_piezoelectric_2018}.

As the structures in the mentioned applications are usually thin, guided elastic waves are of great relevance. These waves propagate dispersively~\cite{auld_acoustic_1990,royer_elastic_2022,langenberg_theoretische_2009}, i.e., their angular frequency~$\omega$ is non-linearly related to their wavenumber~$k$ through the dispersion relation~$\omega(k)$. Of particular interest are solutions where the group velocity component~${\partial \omega / \partial k}$ vanishes while the wavenumber remains finite~\cite{prada_laser-based_2005,clorennec_laser_2006,clorennec_local_2007,balogun_simulation_2007,prada_local_2008,xie_imaging_2019}. In isotropic media, these zero-group-velocity (ZGV) points represent local resonances that are due to the finite thickness of the semi-infinite structure. At sufficiently high frequencies they usually dominate the overall mechanical response of the structure. They are rather simple to excite and measure with contactless laser-ultrasonic techniques. 
ZGV resonances are used in nondestructive evaluation to determine various structural properties such as material parameters, thickness or bonding state with very high precision in a spatially resolved manner~\cite{ces_characterization_2012,grunsteidl_inverse_2016,grunsteidl_determination_2018,mezil_non_2014,ryzy_determining_2023}. 

\begin{figure}[tb]\centering\sffamily\footnotesize%
  % \tikzsetnextfilename{figure_qzgv_decay}
  % \input{tikz/fig_measurement_qzgv_decay}
  \includegraphics{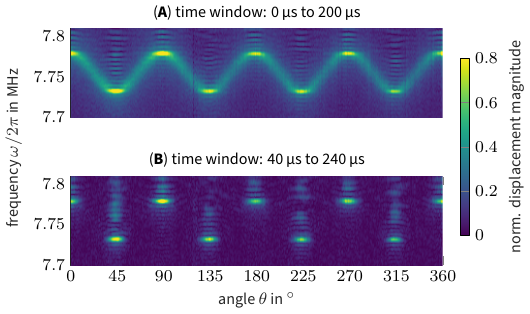}
  \caption{Angular dependence of the ZGV resonance. The surface displacement was acquired at the center of a line source and obtained on two different time windows. The resonance frequency clearly depends on the line orientation with a 90$^\circ$-periodicity. $\theta$ is the angle of the excited wave vector and is counted from the [110] axis. For long times, the resonance remains at $\theta = n \times 45^\circ$, $n \in \mathbb{Z}$, but vanishes elsewhere.
  The measurement data was taken from the experiment detailed in Ref.~\cite{prada_influence_2009}, where (A) has already been reported.}
  \label{fig:resonance_time}
\end{figure}

\begin{figure*}[tb]\centering\sffamily\footnotesize%
  % \tikzset{external/export next=false}
  % \tikzsetnextfilename{wafer_geom}
  % \input{tikz/wafer_geom}
  % \tikzsetnextfilename{fig_waferAndSetup}
  % \input{tikz/fig_waferAndSetup}
  \includegraphics{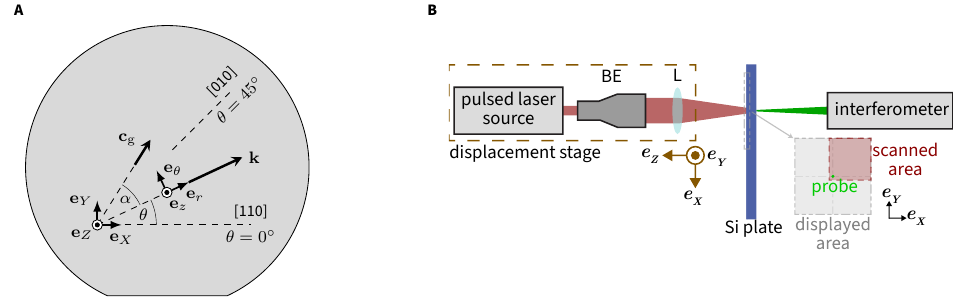}
  \caption{Sample and measurement setup. (\textbf{A}) [001]-cut silicon wafer sample. The ${\dirvec{X}\dirvec{Y}\dirvec{Z}}$ coordinate system is fixed as indicated, while the ${\dirvec{r}\dirvec{\theta}\dirvec{z}}$ system is aligned with the wave propagation direction. The wave vector~$\ten{k}$ is inclined at angle~$\theta$, while the group velocity vector~$\ten{c}_{\mup{g}}$ is at angle $\theta+\alpha$. (\textbf{B}) schematic representation of the laser-ultrasonic measurement setup. BE: Beam Expander, L: Lense.}
  \label{fig:geometry_and_setup}
\end{figure*}

ZGV resonances in anisotropic plates were the object of some theoretical~\cite{kiefer_computing_2023,glushkov_multiple_2021,hussain_lamb_2012} as well as experimental studies~\cite{prada_influence_2009,thelen_laser-excited_2021}. Prada et al.~\cite{prada_influence_2009} revealed the directional dependence of the ZGV resonance frequency in a silicon wafer and we reproduce this result in Fig.~\ref{fig:resonance_time}A (we will show further on that not all of the components really correspond to ZGV resonances). Note that the directional dependence is distinctive of ZGV resonances as this is impossible for common thickness resonances characterized by ${k = 0}$. It is striking that the extremal frequencies in Fig.~\ref{fig:resonance_time} play a special role, in particular, when inspecting later times as depicted in Fig.~\ref{fig:resonance_time}B. Moreover, in Ref.~\cite{prada_influence_2009} the response to the point-source excitation appears to be dominated by the two ZGV frequencies associated to principal directions of the crystal. This fact was confirmed by numerical computations in Ref.~\cite{glushkov_multiple_2021}. However, the described effects remain unexplained in the literature.

The aim of the present work is twofold. Firstly, the special role of ZGV resonances along principal directions of the material is explained. The contributions along other directions are explained by slow waves that propagate parallel to the line source, which means that they exhibit a group-velocity vector that is orthogonal to the wave vector. Secondly, we predict and confirm the formation of complex time-dependent resonance patterns on the surface of the plate after it has been excited by an impulsive point load. We show that this beating pattern can be explained by the interference of eight ZGV modes associated to the material's principal directions.

\section{Results} 
\label{sec:results}

We study transversely propagating waves and ZGV resonances in anisotropic plates. Both effects are confirmed experimentally on a [001]-cut monocrystalline silicon wafer of 524.6\,{\textmu}m thickness, see Fig.~\ref{fig:geometry_and_setup}A. The material's stiffness is of cubic anisotropy (Voigt-notated stiffness $C_{11} = \mathup{165.6\,GPa}$, $C_{12} = \mathup{63.9\,GPa}$, $C_{44} = \mathup{79.5\,GPa}$, density $\rho = \mathup{2330\,kg/m^3}$).
The laser-ultrasonic system depicted in Fig.~\ref{fig:geometry_and_setup}B is used to observe the waves. It consists of a pulsed laser source that is defocused to excite the wavenumber range of interest of up to $\approx$7\,rad/mm and of a laser interferometer that measures the normal surface displacements of the wafer. A two-dimensional scan of the surface is performed by moving the laser source. To speed up the measurement, we only scan one quarter of the wave field and reconstruct the full field by exploiting the sample's material symmetries. Some of the results are bandpass filtered between 7.6\,MHz to 7.8\,MHz in a postprocessing step (indicated explicitly in the figures), as this frequency band contains the phenomena of interest. Further specificities on the measurement setup and the postprocessing can be found in Sec.~\ref{sub:measurement_setup}. Before presenting the experimental results, the underlying theory is developed in the following.

\begin{figure*}[tb]\centering\sffamily\footnotesize%
  % \tikzsetnextfilename{fig_dispersionSurf}
  % \input{tikz/fig_dispersionSurf}
  \includegraphics{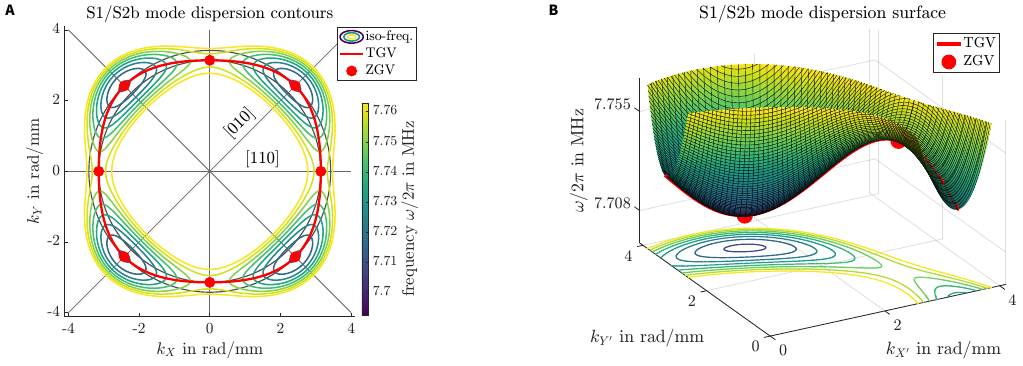}
  \caption{Dispersion surface of the S1/S2b mode close to the minima in frequency. (\textbf{A})~iso-frequency contours in the wave vector plane and (\textbf{B})~surface in the frequency-wave vector space. ZGV: zero-group velocity. TGV: transverse-group velocity. The ZGV points are located at the marked minima and saddle-points (red points). For visualization purposes, (B) has been rotated by 22.5$^\circ$ and cropped to the shown quarter plane.}
  \label{fig:S1S2b_ZGV_surf}
\end{figure*}

\begin{figure*}[tb]\centering\sffamily\footnotesize%
  % \tikzsetnextfilename{fig_dispersionCurves}
  % \input{tikz/fig_dispersionCurves}
  \includegraphics{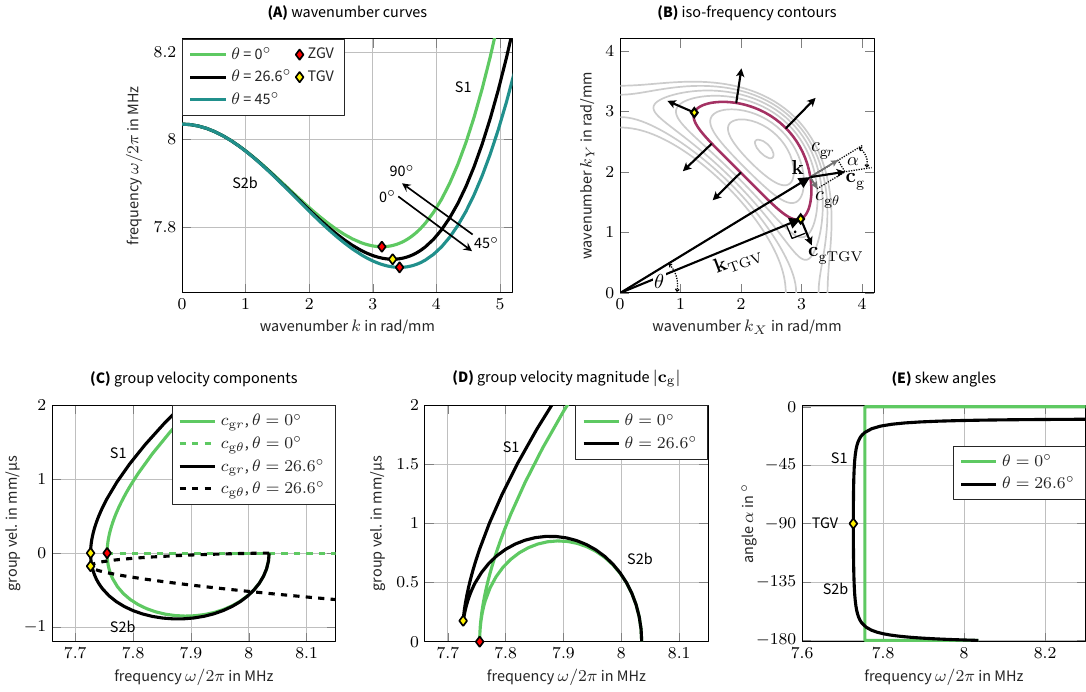}
  \caption{Dispersion of wave vectors and group velocity. (\textbf{A})~frequency vs. wavenumbers. (\textbf{B})~some group velocity vectors sketched on a selected iso-frequency contour. Dispersion curves are cuts along constant $\theta$ of this surface. (\textbf{C})~components $c_{\mup{g}r}$ and $c_{\mup{g}\theta}$ of the group velocity vector vs. frequency, compare also to (B). The transverse power flux due to $c_{\mup{g}\theta}$ does not vanish for TGV waves, while it does vanish at ZGV points. (\textbf{D}) magnitude of the group velocity vector vs. frequency. (\textbf{E})~skew angles~$\alpha$ between the group velocity vectors $\ten{c}_\mup{g}$ and the wave vectors $\ten{k}$ plotted vs. frequency.}
  \label{fig:dispersion}
\end{figure*}

\subsection{Guided waves in a silicon wafer}
\label{sub:guided_waves_in_silicon}
We study guided elastic waves propagating in the wafer. The fixed $\dirvec{X}\dirvec{Y}\dirvec{Z}$ coordinate system is aligned with the [110],[$\bar 1$10] and [001] crystallographic axes, respectively. Moreover, the $\dirvec{r}\dirvec{\theta}\dirvec{z}$ system is oriented with the wave vector, as depicted in Fig.~\ref{fig:geometry_and_setup}A. Taking the point of view of a plane wave, $\dirvec{r}$ denotes the \emph{axial direction}, while $\dirvec{\theta}$ refers to the \emph{transverse direction}. The plate has stress-free surfaces. Furthermore, it is considered to be of infinite lateral dimensions, so that waves reflected from the border can be disregarded.

Guided waves are characterized by the angular frequency~$\omega$, the wave vector~${\ten{k} = k \dirvec{r}(\theta)} = k_X \dirvec{X} + k_Y \dirvec{Y}$, as well as the through-thickness displacement distribution~$\ten{u}(z)$. Only certain combinations of frequency and wave vector can propagate, which is described by the \emph{dispersion relation}~\cite{auld_acoustic_1990,royer_elastic_2022,langenberg_theoretische_2009,kiefer_elastodynamic_2022}. The solutions form surfaces $\omega(k_X, k_Y)$, or equivalently $\omega(k, \theta)$. We use a semi-analytical method to obtain these solutions, i.e., the associated eigenvalue problem is solved numerically. Our implementation is made available as \texttt{GEWtool}~\cite{kiefer_gewtool_2023}. Therewith, we are also able to compute ZGV points (and the transversely propagating waves that will be discussed later on) by employing the numerical methods that we have recently developed for this purpose~\cite{kiefer_computing_2023,plestenjak_gew_2023}. For details on the model and the numerical methods, see Section~\ref{sub:computing_guided_waves_and_power_flux}. 

The computed dispersion surface of the first modes that exhibit ZGV points is shown in Fig.~\ref{fig:S1S2b_ZGV_surf}. Note that due to the cubic anisotropy of silicon, the dispersion surface has an angular periodicity of 90$^\circ$. Furthermore, we observe reflection symmetry across the directions $\theta =~$0$^\circ$ and $\theta =~$45$^\circ$, which correspond to the [110] and [010] crystallographic axes, respectively. Furthermore, the surface exhibits four minima on the $\langle$010$\rangle$ axes and four saddle-points along the $\langle$110$\rangle$ directions.

Often, dispersion curves are plotted along a fixed propagation direction~$\theta$, corresponding to cuts across the dispersion surface ${\omega(k, \theta)}$. The curves in three different propagation directions are depicted in Fig.~\ref{fig:dispersion}A and they differ most in the directions 0$^\circ$ and 45$^\circ$. Although pure Lamb/SH waves do not generally exist in the silicon plate because they are coupled, we label the mode branches consistently to the commonly used notation. In particular, following the notation in Ref.~\cite{prada_local_2008}, the positive-slope branch shown in Fig.~\ref{fig:dispersion}A will be denoted as S1, while the negative-slope section is called S2b. 

The power flux of the waves is proportional to the group velocity~\cite{langenberg_theoretische_2009,langenberg_energy_2010,auld_acoustic_1990,royer_elastic_2022}. The latter is defined as the gradient of $\omega$ with respect to the components of the wave vector, i.e.,
\begin{equation}\label{eq:group_velocity}
   \ten{c}_\mup{g} := \nabla_{\ten{k}} \omega = \frac{\partial \omega}{\partial k} \dirvec{r} + \frac{1}{k} \frac{\partial \omega}{\partial \theta} \dirvec{\theta} \,.
\end{equation} 
It is worth remarking that these vectors are–per definition–orthogonal to the iso-frequency lines drawn in Fig.~\ref{fig:S1S2b_ZGV_surf}A. This is illustrated in Fig.~\ref{fig:dispersion}B, where some group velocity vectors have been sketched for a chosen iso-frequency contour. 

The power flux is discussed by comparing propagation at the principal direction $\theta =~$0$^\circ$ and the nonprincipal direction $\theta =~$26.6$^\circ$. For both directions we compute the group velocity vectors as a function of frequency. The resulting vector components are depicted in Fig.~\ref{fig:dispersion}C, while the magnitude is shown in Fig.~\ref{fig:dispersion}D. Due to the material's reflection symmetry across principal axes, the transverse power flux $c_{\mup{g}\theta}$ is identically zero on principal directions. This is not the case for other propagation directions, as Fig.~\ref{fig:dispersion}C illustrates. Due to this, the corresponding group velocity magnitude in Fig.~\ref{fig:dispersion}D does no longer vanish at the minimal frequency. Furthermore, depending on whether the axial component~$c_{\mup{g}r}$ is positive or negative, we speak of \emph{forward waves} or \emph{backward waves}~\cite{prada_local_2008,shuvalov_backward_2008,bramhavar_negative_2011,legrand_cloaking_2021}, respectively. With respect to a source, a backward wave exhibits outgoing flux but incoming phase fronts. This is the case for the S2b wave. Note that the phenomena investigated hereinafter are strongly related to the existence of such waves.

The transverse group velocity component~$c_{\mup{g}\theta}$ is due to the anisotropy $\partial \omega / \partial \theta$ and causes a transverse power flux. We remark that it is the coupling of Lamb- and SH-polarizations that implies this (usually) nonzero transverse power flux~\cite[Appendix~B]{kiefer_computing_2023}. Due to this transverse component, the overall power flux (or group velocity) is not collinear with the wave vector. The angle between the two is denoted as $\alpha$ and called \emph{steering} or \emph{skew angle} \cite{langenberg_theoretische_2009,chapuis_excitation_2010,karmazin_study_2013}. The frequency-dependent skew angles of the S1/S2b modes are shown in Fig.~\ref{fig:dispersion}E. It is remarkable that they cover almost 180$^\circ$ in a rather narrow frequency range.

\subsection{Zero- and transverse-group-velocity waves: ZGV and TGV}
\label{sub:zgv_resonances_and_tgv_waves}
Resonances appear where the power flux vanishes and such points are denoted as ZGV points. This requires the vanishing of the two components of the  group velocity vector given in (\ref{eq:group_velocity}), i.e., these resonances are associated to stationary points of the dispersion surface of Fig.~\ref{fig:S1S2b_ZGV_surf}. It has been common practice in the literature to plot the dispersion curves for a given propagation direction as in Fig.~\ref{fig:dispersion}A to identify ZGV modes, see Ref~\cite{prada_influence_2009,glushkov_multiple_2021,hussain_lamb_2012}. However, these curves only reveal the axial component $c_{\mup{g}r} = \partial \omega / \partial k$ of the group velocity vector. Thus, identifying ZGV points as extrema of these curves is only valid for isotropic media or for propagation along reflection symmetry planes of any anistroptic material, because in these cases the group-velocity vector is collinear to the wave vector, i.e., $c_{\mup{g}\theta} \sim \partial \omega / \partial \theta \equiv 0$. Hence, while the marked points on the curves of Fig.~\ref{fig:dispersion}A appear to be ZGV points in the conventional sense that $\partial \omega / \partial k = 0$, their transverse power flux might actually be nonzero due to anisotropy. For this reason we distinguish between
\begin{enumerate}
  \item ZGV points: waves with overall zero power flow, i.e., $\partial \omega / \partial k = 0$ and $\partial \omega / \partial \theta = 0$, and
  \item transverse-group-velocity (TGV) waves: waves with zero \emph{axial} power flux, i.e., ${\partial \omega / \partial k = 0}$.
\end{enumerate}

A \emph{ZGV point} is an extremum or saddle point of the dispersion surface depicted in Fig.~\ref{fig:S1S2b_ZGV_surf}. The cubic material exhibits eight isolated ZGV points on the principal directions. The four minima in the $\langle$010$\rangle$ directions are denoted as ZGV1 and occur at $\omega/2\pi =~$7.7079\,MHz and $k =~$3.421\,rad/mm. The ZGV2 are saddle points at $\langle$110$\rangle$ with $\omega/2\pi =~$7.7551\,MHz and $k =~$3.142\,rad/mm. 

The existence of \emph{TGV waves} is evident from the closed iso-frequency contours that enclose a minimum in Fig.~\ref{fig:dispersion}B. These always exhibit points where the wave vector is tangent to the contour and, consequently, orthogonal to the group velocity. The wavenumber and frequency of TGV waves depend on the wave vector orientation~$\theta$ and form the continuous red curve drawn in Fig.~\ref{fig:S1S2b_ZGV_surf}A and Fig.~\ref{fig:S1S2b_ZGV_surf}B. The closed TGV-curve separates the forward wave region (outside) from the backward wave region (inside).

\subsubsection{TGV wave radiated by a line source}
The existence of TGV waves is quite remarkable as their power flux is \emph{orthogonal} to the wave vector. While this fact is obvious from their definition, the evolution of power flux when approaching the TGV-frequency can be assessed in Fig.~\ref{fig:dispersion}E: the skew angles of the S1 and S2b modes both converge to $-$90$^\circ$ when reaching the TGV point. Note that the weaker the coupling between Lamb and SH modes, the sharper we expect the transition towards $-$90$^\circ$ to take place. Ultimately, when these wave families decouple (i.e., on a principal direction or in an isotropic plate), we obtain a discontinuity as observed in Fig.~\ref{fig:dispersion}E for $\theta = 0^\circ$. The combination of orthogonal propagation and wide skew-angle spectrum lead to extraordinary diffraction effects.

\begin{figure*}[tb]\centering\sffamily\footnotesize%
  % \tikzsetnextfilename{fig_qzgv_line_intensity}
  % \input{tikz/fig_qzgv_line_intensity}
  \includegraphics{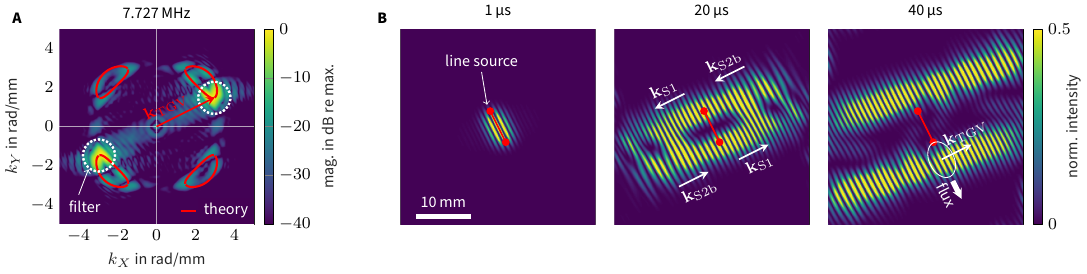}
  \caption{Measured TGV wave field radiated by a synthetic line source inclined 26.6$^\circ$. (\textbf{A})~wavenumber spectrum at the TGV-frequency compared to theory prior to filtering in the wavenumber domain; (\textbf{B})~instantaneous intensity distributions after having applied a Gaussian filter in the wavenumber domain as indicated in~(A). The two wave packets propagate along the line source while the wave vectors are orthogonal. Note that the phases propagate counter-wise in the two pulses, leading to interference where the two overlap. Frequency band: 7.6\,MHz to 7.8\,MHz. Full field reconstructed from one quadrant. (cf. Supplementary Video~1.)}
  \label{fig:tgv_line_intensity}
\end{figure*}

To observe these effects we synthesize a line-source response from the point-source measurements, for details see Sec.~\ref{sub:measurement_setup}. Waves usually radiate in normal direction from the line source and diffract around its edges. However, this is not the case for the TGV wave. This can be observed in Fig.~\ref{fig:tgv_line_intensity}, which presents the response to a 6.4\,mm line source preferentially exciting wave vectors at 26.6$^\circ$ and $-$153.4$^\circ$. The associated wavenumber spectrum at the TGV-frequency is compared to the theoretical wavenumbers in Fig.~\ref{fig:tgv_line_intensity}A. To better observe the TGV wave, we apply a Gaussian filter in the wavenumber plane. Its full width at half maximum is delineated in Fig.~\ref{fig:tgv_line_intensity}A. A subsequent inverse Fourier transform into the spatial domain yields the instantaneous intensity distributions depicted in Fig.~\ref{fig:tgv_line_intensity}B. Two wave packets propagating along the line source can clearly be discerned, i.e., their skew angle is indeed $\pm$90$^\circ$ as expected. From another point of view, this wave ``diffracts'' around the line's edge while maintaining the orientation of the phase fronts aligned with the line source.

A further validation is done by comparing the group velocity of the wave packets to theory. After 40\,{\textmu}s the maxima of the envelops are 6.81\,mm from the source's center, yielding a transverse velocity of $\pm$0.170\,mm/{\textmu}s. This is in very good agreement with the computed transverse group velocity of $-$0.177\,mm/{\textmu}s for the TGV wave at $\theta =~$26.6$^\circ$, as shown in Fig.~\ref{fig:dispersion}C. Note that the wave vectors oriented at $\theta = -$153.4$^\circ$ lead to the wave packet traveling in opposite direction.

Propagation of energy purely orthogonal to the wave vector is a strictly monochromatic process, both in frequency and wavenumber. The neighboring spectral components exhibit a nonzero axial power flux and these waves are responsible for the packets' lateral extend.
Wave packets and their power flux can only be observed when considering a spectrum of finite width. This is ensured by the applied Gaussian filter in Fig.~\ref{fig:tgv_line_intensity}A, which has a half-width of 0.7\,rad/mm. While all spectral components of the wave packets possess very similar wave vectors, their power flux cover a wide range of directions. Indeed, as the power flux is orthogonal to the iso-frequency lines, Fig.~\ref{fig:tgv_line_intensity}A shows that the skew-angle spectrum spans almost 180$^\circ$.

Lastly, it is worth observing that the two wave packet's phases propagate in opposite directions, as indicated in Fig.~\ref{fig:tgv_line_intensity}B. The counter-propagating wave packets lead to quickly varying interference in the regions where the two superpose. It is remarkable that the phase fronts of one packet travel all in the same direction. 
This means that the phases propagate \emph{towards the source} on one side, and \emph{away from the source} on the other side, as indicated by the wave vector arrows in Fig.~\ref{fig:tgv_line_intensity}B. Accordingly, each wave packet clearly consists of the S1 forward wave and the S2b backward wave, separated spatially by the TGV component. This is consistent with the smaller wavelengths that we observe in Fig.~\ref{fig:tgv_line_intensity}B for the S1 components compared to the S2b components. The spatial allocation of modes is also in accordance with Fig.~\ref{fig:dispersion}B as well as Fig.~\ref{fig:dispersion}E.

\subsubsection{Decay of the TGV-wave contributions}
The decay observed for some spectral components in the introductory Fig.~\ref{fig:resonance_time} can now be explained. While the frequency extrema correspond to the ZGV resonances, the remaining spectral components are due to TGV waves. The line source excites the latter all along its length. The TGV waves then propagate along the source, where they are measured at the center point by the interferometer. Due to their propagative nature, the waves leave the detection point after a certain time. As seen in Fig.~\ref{fig:resonance_time}B, the TGV waves disappear completely after about 40\,{\textmu}s, while the expected eight ZGV resonances at $\theta = n \times \text{45}^\circ, n \in \mathbb{Z}$, perdure.

\begin{figure*}[tb]\centering\sffamily\footnotesize%
  % \tikzsetnextfilename{figure_synth}
  % \input{tikz/fig6_resonance_pattern_synth}
  \includegraphics{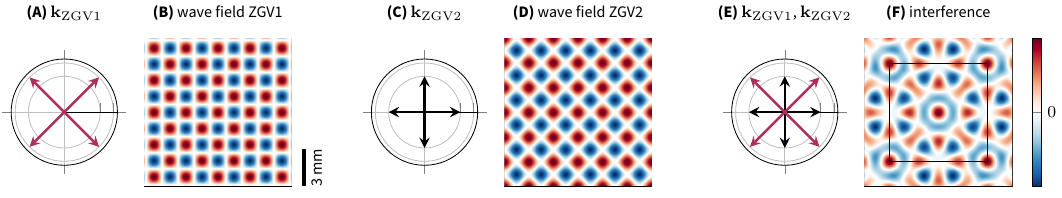}
  \caption{Formation of the S1/S2b-resonance pattern in an infinite silicon plate. (\textbf{A},\textbf{C},\textbf{E})~show the wave vectors (not to scale); while (\textbf{B},\textbf{D},\textbf{F})~depict the corresponding wave fields. The ZGV modes in (\textbf{A},\textbf{B}) and (\textbf{C},\textbf{D})~are characterized by close but different wavenumbers $|\ten{k}|$ and frequencies $\omega$. These two standing wave fields interfere to form a non-stationary, time-dependent beating pattern as shown for example in~(\textbf{F}) when the phase shift between the two is zero. (cf. Supplementary Video~2.)}
  \label{fig:resonance_pattern_synth}
\end{figure*}

The time for the TGV waves to escape the measurement point depends both on the length of the line source and the wave's group velocity. Compared to other wave components, the TGV wave exhibits a very low group velocity, as can well be appreciated in Fig.~\ref{fig:dispersion}D. Therefore, the TGV waves are detected for a relatively long time, resulting in a well pronounced spectral peak in Fig.~\ref{fig:resonance_time}A. This explains why they play an important role in the presented measurements. Note that the group velocity tends to zero as the line orientation tends to one of the material's principal axes. Hence, the closer the line orientation is to a ZGV point, the more pronounced the corresponding spectral peak will be.

\subsection{Resonance pattern of ZGV modes}
\label{sub:resonance_pattern}

The previous results demonstrate that time acts as a filter in the wave vector-frequency domain. After sufficient time, only the ZGV resonances remain in the finite spatial region of observation. As a consequence, a resonance pattern develops in the silicon crystal plate. We first lay out the theory explaining this resonance and subsequently we present direct measurements thereof. 

\subsubsection{Theory of ZGV resonances}
\label{ssub:theory_of_the_zgv_resonance}

For the frequency region of interest, the resonance pattern is explained by the interference of the eight ZGV resonances sketched in Fig.~\ref{fig:resonance_pattern_synth}A and Fig.~\ref{fig:resonance_pattern_synth}C. 
The interference of the four wave vectors pertaining to a given ZGV mode leads to the periodic square wave pattern depicted in Fig.~\ref{fig:resonance_pattern_synth}B and Fig.~\ref{fig:resonance_pattern_synth}D. Note that for each wave vector $\ten{k}$, there exists a counter-propagating wave $-\ten{k}$. As a consequence, the square patterns are standing wave fields~\cite{prada_local_2008}. It is notable that this standing wave field is not due to the edges of the plate but actually emerges in the \emph{infinite plate}. 

It is pertinent to discuss differences between the isotropic and anisotropic cases. While the isotropic ZGV resonance consist of a continuous wave vector spectrum, i.e., it exhibits wave vectors in all possible directions, the ZGV resonances in an anisotropic medium consist of a finite set of wave vectors. Accordingly, the ZGV field due to a point load on an isotropic plate is a cylindrical standing wave~\cite{xie_imaging_2019}. In particular, this means that the nodal curves seen on the plate's surface are closed circles around the source that never cross. In contrast to this, the field of the ZGV1 or ZGV2 resonances are ``checkerboard standing waves'' as shown in Fig.~\ref{fig:resonance_pattern_synth}B and \ref{fig:resonance_pattern_synth}D. Hence, anisotropy leads to straight, open and crossing nodal lines. 

The ZGV1 and ZGV2 waves interfere to form a resonance pattern as depicted in Fig.~\ref{fig:resonance_pattern_synth}F. The nodal curves form ``bubbles'' of different shapes that are no longer square. As each of the wave components is flux free, no energy is propagated. Nonetheless, this is no longer a standing wave field because the ZGV1 and ZGV2 components are at different frequencies. Instead, the ``bubbles'' move radially out or into the center of the pattern while changing their shape (see Supplementary Video~2). Note that they shortly stop moving when the two ZGV components are in phase or in paraphase. For the field synthesis shown in Fig.~\ref{fig:resonance_pattern_synth}F we arbitrarily assumed identical amplitudes and a zero phase shift between ZGV1 and ZGV2. Due to the similarity of the displacement eigenfunctions $\ten{u}_\mup{ZGV1}(z)$ and $\ten{u}_\mup{ZGV2}(z)$, both modes are excited similarly by the point source. For the same reason, we expect both ZGVs to be in phase at the moment of generation by the pulse.

The phase shift between ZGV1 and ZGV2 is modulated in time due to their slightly different frequencies. In other words, temporal beating is expected as a consequence of their superposition. Hence, the instantaneously observed phase shift loops $2\pi$ in a period of 
\begin{equation}\label{eq:beating_period}
  \Delta T = \frac{2\pi}{\omega_\mup{ZGV2} - \omega_\mup{ZGV1}} = \text{{21.2\,\textmu}s}\,.
\end{equation}

\subsubsection{Measurement of ZGV resonances}
\label{ssub:measurement_of_the_zgv_resonance}

\begin{figure*}[tb]\centering\sffamily\footnotesize%
  % \tikzsetnextfilename{figure_meas_time}
  % \input{tikz/fig_meas_time}
  \includegraphics{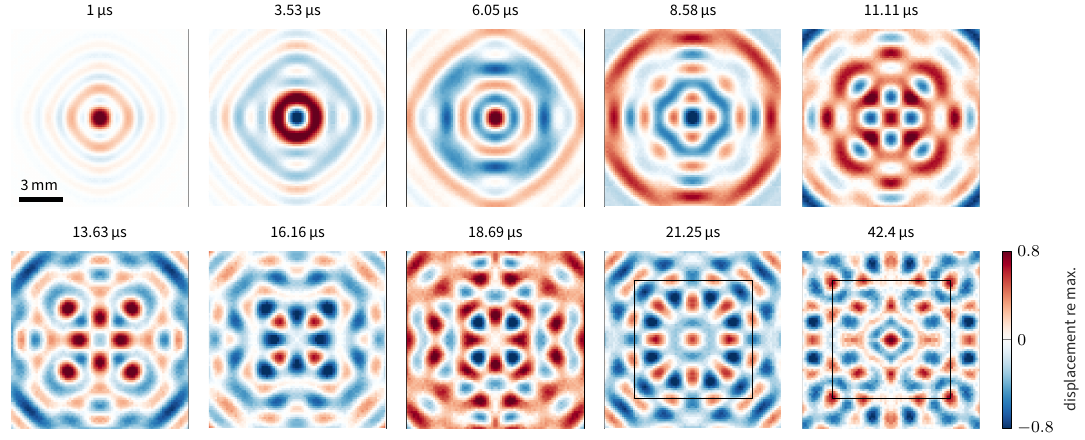}
  \caption{Measured wave fields at selected times. A resonance pattern forms due to the superposition of the eight plane waves corresponding to ZGV resonances in the single-crystal silicon wafer. The pattern forms naturally over time as only the ZGV resonances remain in the spatial window close to the point source. The inner region of {4\,mm $\times$ 4\,mm} is marked for reference. Frequency band: 7.6\,MHz to 7.8\,MHz. Full field reconstructed from one quadrant. (cf. Supplementary Video~3 and~4.)}
  \label{fig:resonance_pattern}
\end{figure*}

The described resonance pattern is directly observable by measuring the wave field excited by a point-like source. As the S1/S2b resonance is the one that is most strongly excited by our system, the theoretically expected moving resonance pattern can be discerned in the raw data without postprocessing, see Supplementary Video~4. To obtain a clean result, other resonances as well as reflections from the border of the plate should be avoided.
We achieve this with the mentioned frequency bandpass filter in combination with the large source that excites low wavenumbers, see Sec.~\ref{sub:measurement_setup} for details. The thus observed frequency-wavenumber range contains only slow modes in the vicinity of the ZGV resonances. The filtered displacement fields are depicted in Fig.~\ref{fig:resonance_pattern} for selected time instants.

Propagating waves dominate the picture at first and their phase fronts (nodal curves) encircle the source. But a nodal curve pattern with the mentioned ``bubbles'' forms over time, see Fig.~\ref{fig:resonance_pattern} and Supplementary Video~3. The pattern develops slowly from the source (in the center) outwards as the propagative waves leave the region of excitation. When the pattern becomes first visible, the two ZGV resonances are not in phase and the ``bubbles'' move in and out on the $\langle$110$\rangle$ and $\langle$100$\rangle$ axes as expected from the theory. 

After one beating period $\Delta T$, the two ZGVs should be in-phase for the first time and, hence, we expect a pattern similar to Fig.~\ref{fig:resonance_pattern_synth}F. At 21.25\,{\textmu}s the pattern in the inner area marked by the square in Fig.~\ref{fig:resonance_pattern} resembles the corresponding region in Fig.~\ref{fig:resonance_pattern_synth}F. Outside this region, the contribution of propagating waves is still too high. Nonetheless, after two beating periods, at 42.4{\textmu}s, the non-ZGV waves propagated and we can now recognize a fully developed ZGV resonance pattern in the region of observation. Note, for example, the ``eight-pointed star'' centered at the center and at the edges of the marked inner region. Discrepancies to Fig.~\ref{fig:resonance_pattern_synth}F can mostly be attributed to temporal sampling of the measurement.

\begin{figure}[tb]\centering\sffamily\footnotesize%
  % \tikzsetnextfilename{figure_measured_fields_ZGV}
  % \input{tikz/fig_measured_fields_ZGV}
  \includegraphics{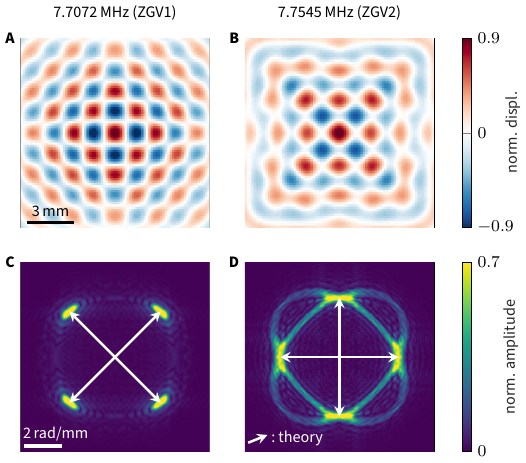}
  \caption{Measured displacement fields at the two ZGV frequencies. (\textbf{A},\textbf{B})~harmonic field at arbitrary phase in the physical $X$-$Y$-space obtained by a temporal Fourier transform. (\textbf{C},\textbf{D})~spectral magnitude in the reciprocal $k_X$-$k_Y$-space obtained by a spatio-temporal Fourier transform. The numerically computed ZGV wave vectors are indicated by arrows. Full field reconstructed from one quadrant.}
  \label{fig:measurement_fields}
\end{figure}

A quantitative validation is performed by comparing measurement and theory in the spectral domain. First, a time-Fourier transform yields the measured wave fields at the ZGV frequencies. The result is depicted in Fig.~\ref{fig:measurement_fields}A and Fig.~\ref{fig:measurement_fields}B and both confirm the wave fields expected in Figs.~\ref{fig:resonance_pattern_synth}B and \ref{fig:resonance_pattern_synth}D. Second, we additionally perform a spatial 2D Fourier transform into the wave vector-frequency domain. The obtained spectral amplitudes are displayed in Fig.~\ref{fig:measurement_fields}C and Fig.~\ref{fig:measurement_fields}D together with the ZGV wave vectors expected from theory. We observe that the energy confines closely to the regions predicted by the computed wave vectors. While the wave field at $\omega_\mup{ZGV1}$ consists almost purely of the ZGV resonances, propagating modes exist at $\omega_\mup{ZGV2}$ and are also excited. This is because $\omega_\mup{ZGV2}$ corresponds to the saddle-points and the corresponding iso-frequency contours are two curves that cross/touch at the ZGV2-points. In contrast to this, the iso-frequency contours at $\omega_\mup{ZGV1}$ reduce to the isolated ZGV1-points. Note that spatio-temporal gating can be used to remove the propagating waves of Fig.~\ref{fig:measurement_fields}D, as was done in Fig.~\ref{fig:resonance_time}. Overall, this confirms that the resonances observed in an infinite plate of anisotropic elasticity are formed by the superposition of a discrete set of ZGV modes.

\section{Discussion}
\label{sec:Discussion}
The existence of isolated critical points in the S1/S2b mode dispersion surface entails two distinct and extraordinary effects: transverse wave propagation (TGV waves) and beating ZGV resonances. We presented the underlying theory as well as experimental evidence of both effects. TGV waves are characterized by a power flux that is orthogonal to their wave vector. They propagate along a line source that excites them. For this reason they manifest as a peak in the spectrum measured at a chosen point of the line source.

ZGV resonances are already used very successfully for ultrasonic nondestructive testing and material characterization. With the developed physical understanding, we expect similar procedures to become attractive for novel microelectromechanical systems, in particular sensors. The present work overcomes one of the major difficulties in designing such systems: properly accounting for anisotropy, such as is encountered for silicon.

It is evident form the theory we presented that the ZGV resonances directly reveal information about the principal axes of the material, independent of the actual orientation of the sample. Measuring the ZGV1 and ZGV2 resonance frequencies as well as the frequency of a regular thickness resonance is sufficient to fully characterize a material of cubic symmetry. These frequencies are readily obtained from the recorded normal surface displacement after a single pulsed excitation because they all manifest as sharp peaks in the spectrum~\cite{prada_influence_2009}. This avoids the technically complicated scan of the wave field, as needs to be done for widespread techniques based on guided waves. Overall, this could enable nondestructive, contactless and real-time characterization of complex materials. 

The present contribution restricted to ZGV resonances of the S1/S2b modes. However, it is well known that other ZGV modes exist at higher-frequencies. They are substantially weaker and were frequency-filtered in this work. Note that anisotropic plates can also support multiple ZGV resonances pertaining to the same modes~\cite{kiefer_computing_2023}. In this case, not only local minima, but also local maxima can appear in the dispersion surface. The analysis and mechanics presented here apply analogously. 

The discussed phenomena are very particular to two-dimensional (and partly also three-dimensional) wave propagation in anisotropic media. Analogous phenomena should be expected when the material pertains to a different anisotropy class than the one studied here. Moreover, similar effects are expected in phononic crystals or meta-materials, as dispersion surfaces with similar features haven been reported for these materials~\cite{bossart_extreme_2023,maznev_mapping_2011,kaina_negative_2015,yves_crystalline_2017}. But note that the critical points (ZGV points) that we studied here exist even without periodicity, i.e., in a homogeneous anisotropic plate. Lastly, hyperbolic polaritons share many of the wave propagation features described here, including power flux orthogonal to the wave vector~\cite{zhang_ultrafast_2023}.

\section{Methods}
\label{sec:methods}

The procedure that we use to compute guided elastic waves is sketched in Subsec.~\ref{sub:computing_guided_waves_and_power_flux}. The experiment is detailed in Subsec.~\ref{sub:measurement_setup}.

\subsection{Computing guided waves and power flux}
\label{sub:computing_guided_waves_and_power_flux}
The waves propagate in a plate that is assumed to be infinite in the $X$-$Y$-plane, see Fig.~\ref{fig:geometry_and_setup}A.
Its material is of anisotropic stiffness~$\ten{c}$ and mass density~$\rho$. The plane waves of wavenumber~$k$ shall be harmonic with angular frequency~$\omega$. Hence, their wave field is of the form 
\begin{equation}
  \ten{\tilde{u}}(r,\theta,z,t) = \ten{u}(z) \e^{\iu k r} \e^{-\iu \omega t} \,.
\end{equation}
We proceed by arbitrarily fixing the propagation vector $\ten{k} = k \dirvec{r}$ and computing the corresponding frequency~$\omega$ and modal displacements~$\ten{u}(z)$. These are obtained as the solutions of a differential eigenvalue problem~\cite{adamou_spectral_2004,kiefer_elastodynamic_2022,kiefer_computing_2023}. $\omega$ and $\ten{u}(z)$ represent the eigenvalues and the eigenfunctions, respectively. For a concise derivation of the concrete problem formulation see Ref.~\cite{kiefer_computing_2023}. 

We compute solutions with a \emph{semi-analytical procedure} that consists in two steps: (i) discretize the differential eigenvalue problem and (ii) solve the resulting algebraic eigenvalue problem numerically. Many variants of this procedure have been discussed in the literature~\cite{adamou_spectral_2004,gravenkamp_numerical_2012,finnveden_evaluation_2004}. A concise implementation based on spectral collocation is \texttt{GEW dispersion script}~\cite{kiefer_gew_2022}. For the current work, we perform the discretization using the \emph{spectral element method}, i.e., one finite element of high polynomial order~\cite{gravenkamp_high-order_2021}. This method leads to a regular Hermitian eigenvalue problem~\cite{gravenkamp_notes_2023}, which allows us to reliably compute the ZGV and TGV points in general elastic waveguides~\cite{kiefer_computing_2023,plestenjak_gew_2023}. We make our implementations available under the name of \texttt{GEWtool}~\cite{kiefer_gewtool_2023} and it includes all required post-processing methods to reproduce the results of this contribution. In particular, the script \texttt{dispersionSurface\_silicon\_ZGV.m} in the \texttt{examples} directory produces Fig.~\ref{fig:S1S2b_ZGV_surf}A. 
%The computational methods to compute ZGV points are included in \texttt{GEWtool} and are also available separately~\cite{}.

The coupling and decoupling of wave families is of importance~\cite{kiefer_elastodynamic_2022,kiefer_computing_2023,hernando_quintanilla_symmetry_2017,gravenkamp_notes_2023}. Symmetric and anti-symmetric waves decouple due to the plate's symmetry across its mid-plane. We only compute the symmetric waves by modeling the top half of the plate and fixing the $u_z$ displacement component at the mid-plane.
Moreover, the shear-horizontal (SH) polarization ($u_\theta$) only decouples form the Lamb polarization ($u_r, u_z$) for $\theta = n \times \text{45}^\circ, n \in \mathbb{Z}$. For this reason, we always compute the \emph{fully-coupled waves}, meaning that all three displacement components are accounted for in the displacement vector $\ten{u}(z)$. 

After computing guided wave solutions, their power flux and group velocity can be computed in a post-processing step. To this end, we exploit the fact that in nondissipative waveguides (more precisely: real-valued $\ten{k}$ and $\omega$) the group velocity is equal to the \emph{energy velocity}~\cite{langenberg_energy_2010,auld_acoustic_1990,biot_general_1957} and is given by
\begin{equation}\label{eq:energy_vel}
  \ten{c}_\mup{g} = \ten{c}_\mup{e} = \frac{\int \ten{p} \diff z}{\int  \mathcal{H} \diff z} \,.
\end{equation}
It is defined through the power flux density vector~$\ten{p}$ and the average total stored energy density~$\mathcal{H}$. Using the particle velocity~$\ten{v} = -\iu\omega \ten{u}$ and stress $\ten{T} = { \ten{c}:\nabla \ten{u} } = {\ten{c} : (\iu k \dirvec{r} + \dirvec{z} \partial_z) \ten{u} }$, the power flux density can be computed as 
\begin{align}
  \ten{p} &= -\frac{1}{2} \Re\{ \ten{v}^* \cdot \ten{T} \} \nonumber\\
  \label{eq:power_flux}
  &= -\frac{1}{2} \Re\{ \iu\omega \ten{u}^* \cdot \ten{c} : (\iu k \dirvec{r} + \dirvec{z} \partial_z) \ten{u} \} \,,
\end{align}
where ``$^*$'' denotes complex conjugation, ``$\Re$'' the real-part operator and $\partial_z = \partial/\partial z$. Furthermore, due to equipartition of energy~\cite{langenberg_theoretische_2009}, we can compute the average total stored energy density by
\begin{equation}\label{eq:stored_energy}
  \mathcal{H} = \frac{1}{2} \rho \omega^2 \ten{u}^* \cdot \ten{u} \,.
\end{equation}

Our previously outlined procedure to solve for guided waves yields all quantities required to compute equation (\ref{eq:power_flux}) and equation (\ref{eq:stored_energy}). The differentiation in equation (\ref{eq:power_flux}) and the integration in equation (\ref{eq:energy_vel}) can be performed numerically. In this way, we can compute the energy velocity of each point of the dispersion curves independently. Most importantly, proceeding in this way explicitly provides both the axial and the transverse components of the energy velocity vector. The $z$-component is identically zero due to the power-flux-free surfaces of the plate.

Lastly, as the wave vector is $\ten{k} = k \dirvec{r}$, the \emph{skew angle}~$\alpha$ can be computed from the energy velocity vector as 
\begin{equation}
  \alpha = 
  \begin{cases}
    \arctan (c_{\mup{e}\theta}/c_{\mup{e}r}) & \text{for } c_{\mup{e}r} \ge 0 \,,\\
    \arctan (c_{\mup{e}\theta}/c_{\mup{e}r}) - \pi & \text{otherwise} \,.
  \end{cases}
\end{equation}

\subsection{Measurement setup}
\label{sub:measurement_setup}
Measurements are achieved all-optically and are presented schematically in Fig.~\ref{fig:geometry_and_setup}B. Guided waves are generated with a Q-switched  Nd:YAG (yttrium aluminum garnet) laser (Quantel Laser, France, Centurion, 1064\,nm optical wavelength, 100\,Hz repetition rate) that delivers 10\,ns pulses of 9\,mJ. The laser output beam is expanded and then focused onto the silicon plate with a lens (100\,mm focal length). The focal spot is kept rather large (beam width {$\approx$1\,mm}), thus exciting wavenumbers up to {$\approx$7\,rad/mm}. This favors the generation of the first ZGV resonance while limiting the generation of the fundamental guided modes in the frequency range of interest~\cite{bruno16}. The wafer (525\,{\textmu}m nominal thickness, 524.6\,{\textmu}m measured thickness, 125\,mm diameter) has a 100\,nm aluminum coating on the excitation side, which reduces the optical penetration depth. However, similar results were obtained when exciting on the side without the coating. The coating is thin enough not to affect the elastic waves in the silicon plate. 

The normal surface displacement is detected on the opposite side with a heterodyne interferometer of 532\,nm optical wavelength with a power of 100\,mW and a focal spot of {$\approx$50\,\textmu m}. A hardware high-pass filter with 1.25~MHz cut-off frequency is used to avoid saturation of the interferometer due to the large low-frequency displacements of the A0 mode. Signals are recorded with 100\,MHz sampling rate by an oscilloscope connected to a computer. Each signal is averaged 128 times. The {25.05\,mm\,$\times$\,25.05\,mm} scan is performed by moving the excitation unit with a two-axes translation stage along $\mathbf{e}_X$ and $\mathbf{e}_Y$ with a 150\,{\textmu m} pitch. Note that scanning the field is time consuming. For this reason we exploited the cubic symmetry of the material and scanned only one quadrant around the source, as indicated in the inset of Fig.~\ref{fig:geometry_and_setup}B. The full fields have been reconstructed by an appropriate symmetrization.

While the measurements directly provide the point-source response, the line-source response shown in Fig.~\ref{fig:tgv_line_intensity} is synthesized a posteriori. To this end, we superpose 20 shifted point-source responses. The synthetic line source exciting wave vectors at $\theta =~$26.6$^\circ$ is obtained by shifting two pitches vertically for every horizontal one.

Fast waves outside the ZGV region are also excited and their reflections from the border of the plate disturb the long-time observation of the slow TGV waves in Fig.~\ref{fig:tgv_line_intensity} as well as the resonance patterns in Fig.~\ref{fig:resonance_pattern}. To avoid this, we use a band-pass filtered between 7.6\,MHz and 7.8\,MHz for these figures. Note that due to the large source size, large wavenumbers are not observed. The remaining frequency-wavenumber range contains only the slow modes close to the ZGV points.

\appendix
% \bibliographystyle{Science}
% \bibliographystyle{unsrt}
% \bibliography{ZGVpartial}
\printbibliography

@article{zhang_ultrafast_2023,
	title        = {Ultrafast anisotropic dynamics of hyperbolic nanolight pulse propagation},
	author       = {Zhang, Xin and Yan, Qizhi and Ma, Weiliang and Zhang, Tianning and Yang, Xiaosheng and Zhang, Xinliang and Li, Peining},
	year         = 2023,
	month        = aug,
	journal      = {Sci. Adv.},
	volume       = 9,
	number       = 34,
	pages        = {eadi4407},
	doi          = {10.1126/sciadv.adi4407}
}

@software{kiefer_gewtool_2023,
	title        = {{GEWtool}},
	author       = {Kiefer, Daniel A.},
	year         = 2023,
	month        = nov,
	doi          = {10.5281/zenodo.10114244},
	url          = {https://github.com/dakiefer/GEWtool},
	copyright    = {ISC License},
	version      = {1.0.0},
	note         = {{DOI}: \href{https://dx.doi.org/10.5281/zenodo.10114244}{10.5281/zenodo.10114244}}
}

@article{gravenkamp_notes_2023,
	title        = {Notes on osculations and mode tracing in semi-analytical waveguide modeling},
	author       = {Gravenkamp, Hauke and Plestenjak, Bor and Kiefer, Daniel A.},
	year         = 2023,
	month        = jul,
	journal      = {Ultrasonics},
	pages        = 107112,
	doi          = {10.1016/j.ultras.2023.107112},
	issn         = {0041-624X},
}

@article{yves_crystalline_2017,
	title        = {Crystalline {Soda} {Can} {Metamaterial} exhibiting {Graphene}-like {Dispersion} at subwavelength scale},
	author       = {Yves, Simon and Lemoult, Fabrice and Fink, Mathias and Lerosey, Geoffroy},
	year         = 2017,
	month        = nov,
	journal      = {Sci. Rep.},
	volume       = 7,
	number       = 1,
	pages        = 15359,
	doi          = {10.1038/s41598-017-15335-3},
	issn         = {2045-2322},
	copyright    = {2017 The Author(s)},
}

@article{kaina_negative_2015,
	title        = {Negative refractive index and acoustic superlens from multiple scattering in single negative metamaterials},
	author       = {Kaina, Nadège and Lemoult, Fabrice and Fink, Mathias and Lerosey, Geoffroy},
	year         = 2015,
	month        = sep,
	journal      = {Nature},
	volume       = 525,
	number       = 7567,
	pages        = {77--81},
	doi          = {10.1038/nature14678},
	issn         = {1476-4687},
	copyright    = {2015 Springer Nature Limited},
}

@article{bossart_extreme_2023,
	title        = {Extreme {Spatial} {Dispersion} in {Nonlocally} {Resonant} {Elastic} {Metamaterials}},
	author       = {Bossart, Aleksi and Fleury, Romain},
	year         = 2023,
	month        = may,
	journal      = {Phys. Rev. Lett.},
	volume       = 130,
	number       = 20,
	pages        = 207201,
	doi          = {10.1103/PhysRevLett.130.207201},
	issn         = {0031-9007, 1079-7114},
}

@article{bruno16,
	title        = {Laser beam shaping for enhanced Zero-Group Velocity Lamb modes generation},
	author       = {Bruno, Fran{\c c}ois and Laurent, J{\'e}r{\^o}me and Jehanno, Paul and Royer, Daniel and Prada, Claire},
	year         = 2016,
	journal      = {J. Acoust. Soc. Am.},
	volume       = 140,
	number       = 4,
	pages        = {2829--2838},
	doi          = {10.1121/1.4965291},
	issn         = {0001-4966},
}

@article{adamou_spectral_2004,
	title        = {Spectral methods for modelling guided waves in elastic media},
	author       = {Adamou, A. T. I. and Craster, R. V.},
	year         = 2004,
	month        = sep,
	journal      = {J. Acoust. Soc. Am.},
	volume       = 116,
	number       = 3,
	pages        = {1524--1535},
	doi          = {10.1121/1.1777871},
	issn         = {0001-4966}
}

@article{hernando_quintanilla_symmetry_2017,
	title        = {The symmetry and coupling properties of solutions in general anisotropic multilayer waveguides},
	author       = {Hernando Quintanilla, F. and Lowe, M. J. S. and Craster, R. V.},
	year         = 2017,
	month        = jan,
	journal      = {J. Acoust. Soc. Am.},
	volume       = 141,
	number       = 1,
	pages        = {406--418},
	doi          = {10.1121/1.4973543},
	issn         = {0001-4966}
}

@book{auld_acoustic_1990,
	title        = {Acoustic {Fields} and {Waves} in {Solids}},
	author       = {Auld, B. A.},
	year         = 1990,
	publisher    = {Krieger Publishing Company},
	address      = {Malabar},
	volume       = 2,
	isbn         = {978-0-89874-783-6},
	edition      = 2,
}

@book{langenberg_theoretische_2009,
	title        = {Theoretische {Grundlagen} der zerstörungsfreien {Materialprüfung} mit {Ultraschall}},
	author       = {Langenberg, Karl-Jörg and Marklein, René and Mayer, Klaus},
	year         = 2009,
	publisher    = {De Gruyter Oldenbourg},
	address      = {Munich},
	isbn         = {978-3-486-58881-1},
	edition      = 1,
}

@article{prada_local_2008,
	title        = {Local vibration of an elastic plate and zero-group velocity {Lamb} modes},
	author       = {Prada, Claire and Clorennec, Dominique and Royer, Daniel},
	year         = 2008,
	month        = jul,
	journal      = {J. Acoust. Soc. Am.},
	volume       = 124,
	number       = 1,
	pages        = {203--212},
	doi          = {10.1121/1.2918543},
	issn         = {0001-4966}
}

@inproceedings{langenberg_energy_2010,
	title        = {Energy vs. group velocity for elastic waves in homogeneous anisotropic solid media},
	author       = {Langenberg, Karl J. and Marklein, Rene and Mayer, Klaus},
	year         = 2010,
	month        = aug,
	booktitle    = {2010 {URSI} {International} {Symposium} on {Electromagnetic} {Theory}},
	publisher    = {IEEE},
	address      = {Berlin, Germany},
	pages        = {733--736},
	doi          = {10.1109/URSI-EMTS.2010.5637253},
	isbn         = {978-1-4244-5155-5}
}

@article{finnveden_evaluation_2004,
	title        = {Evaluation of modal density and group velocity by a finite element method},
	author       = {Finnveden, S.},
	year         = 2004,
	month        = may,
	journal      = {J. Sound Vib.},
	volume       = 273,
	number       = 1,
	pages        = {51--75},
	doi          = {10.1016/j.jsv.2003.04.004},
	issn         = {0022-460X},
}

@article{prada_laser-based_2005,
	title        = {Laser-based ultrasonic generation and detection of zero-group velocity {Lamb} waves in thin plates},
	author       = {Prada, C. and Balogun, O. and Murray, T. W.},
	year         = 2005,
	month        = nov,
	journal      = {Appl. Phys. Lett.},
	volume       = 87,
	number       = 19,
	pages        = 194109,
	doi          = {10.1063/1.2128063},
	issn         = {0003-6951, 1077-3118},
}

@article{clorennec_laser_2006,
	title        = {Laser impulse generation and interferometer detection of zero group velocity {Lamb} mode resonance},
	author       = {Clorennec, Dominique and Prada, Claire and Royer, Daniel and Murray, Todd W.},
	year         = 2006,
	month        = jul,
	journal      = {Appl. Phys. Lett.},
	volume       = 89,
	number       = 2,
	pages        = {024101},
	doi          = {10.1063/1.2220010},
	issn         = {0003-6951, 1077-3118},
}

@article{balogun_simulation_2007,
	title        = {Simulation and measurement of the optical excitation of the {S1} zero group velocity {Lamb} wave resonance in plates},
	author       = {Balogun, O. and Murray, T. W. and Prada, C.},
	year         = 2007,
	month        = sep,
	journal      = {J. Appl. Phys.},
	volume       = 102,
	number       = 6,
	pages        = {064914},
	doi          = {10.1063/1.2784031},
	issn         = {0021-8979, 1089-7550},
}

@article{thelen_laser-excited_2021,
	title        = {Laser-excited elastic guided waves reveal the complex mechanics of nanoporous silicon},
	author       = {Thelen, Marc and Bochud, Nicolas and Brinker, Manuel and Prada, Claire and Huber, Patrick},
	year         = 2021,
	month        = jun,
	journal      = {Nat. Commun.},
	volume       = 12,
	number       = 1,
	pages        = 3597,
	doi          = {10.1038/s41467-021-23398-0},
	issn         = {2041-1723},
	copyright    = {2021 The Author(s)},
}

@article{mezil_non_2014,
	title        = {Non contact probing of interfacial stiffnesses between two plates by zero-group velocity {Lamb} modes},
	author       = {Mezil, Sylvain and Laurent, Jérôme and Royer, Daniel and Prada, Claire},
	year         = 2014,
	month        = jul,
	journal      = {Appl. Phys. Lett.},
	volume       = 105,
	number       = 2,
	pages        = {021605},
	doi          = {10.1063/1.4890110},
	issn         = {0003-6951, 1077-3118},
}

@article{xie_imaging_2019,
	title        = {Imaging gigahertz zero-group-velocity {Lamb} waves},
	author       = {Xie, Qingnan and Mezil, Sylvain and Otsuka, Paul H. and Tomoda, Motonobu and Laurent, Jérôme and Matsuda, Osamu and Shen, Zhonghua and Wright, Oliver B.},
	year         = 2019,
	month        = dec,
	journal      = {Nat. Commun.},
	volume       = 10,
	number       = 1,
	pages        = 2228,
	doi          = {10.1038/s41467-019-10085-4},
	issn         = {2041-1723},
}

@article{grunsteidl_inverse_2016,
	title        = {Inverse characterization of plates using zero group velocity {Lamb} modes},
	author       = {Grünsteidl, Clemens and Murray, Todd W. and Berer, Thomas and Veres, István A.},
	year         = 2016,
	month        = feb,
	journal      = {Ultrasonics},
	volume       = 65,
	pages        = {1--4},
	doi          = {10.1016/j.ultras.2015.10.015},
	issn         = {0041-624X},
}

@book{kiefer_elastodynamic_2022,
	title        = {Elastodynamic quasi-guided waves for transit-time ultrasonic flow metering},
	author       = {Kiefer, Daniel A.},
	year         = 2022,
	month        = jul,
	publisher    = {FAU University Press},
	address      = {Erlangen},
	series       = {{FAU} {Forschungen}, {Reihe} {B}, {Medizin}, {Naturwissenschaft}, {Technik}},
	number       = 42,
	doi          = {10.25593/978-3-96147-550-6},
	isbn         = {978-3-96147-549-0}
}

@article{shuvalov_backward_2008,
	title        = {On the backward {Lamb} waves near thickness resonances in anisotropic plates},
	author       = {Shuvalov, A. L. and Poncelet, O.},
	year         = 2008,
	month        = jun,
	journal      = {Int. J. Solids Struct.},
	volume       = 45,
	number       = 11,
	pages        = {3430--3448},
	doi          = {10.1016/j.ijsolstr.2008.02.004},
	issn         = {0020-7683},
}

@article{glushkov_multiple_2021,
	title        = {Multiple zero-group velocity resonances in elastic layered structures},
	author       = {Glushkov, E. V. and Glushkova, N. V.},
	year         = 2021,
	month        = may,
	journal      = {J. Sound Vib.},
	volume       = 500,
	pages        = 116023,
	doi          = {10.1016/j.jsv.2021.116023},
	issn         = {0022-460X},
}

@article{hussain_lamb_2012,
	title        = {Lamb modes with multiple zero-group velocity points in an orthotropic plate},
	author       = {Hussain, Takasar and Ahmad, Faiz},
	year         = 2012,
	month        = aug,
	journal      = {J. Acoust. Soc. Am.},
	volume       = 132,
	number       = 2,
	pages        = {641--645},
	doi          = {10.1121/1.4730891},
	issn         = {0001-4966},
}

@book{royer_elastic_2022,
	title        = {Elastic {Waves} in {Solids} 1: {Propagation}},
	shorttitle   = {Elastic {Waves} in {Solids} 1},
	author       = {Royer, Daniel and Valier-Brasier, Tony},
	year         = 2022,
	month        = apr,
	publisher    = {ISTE and John Wiley \& Sons},
	address      = {London},
	isbn         = {978-1-78630-814-6},
}

@article{ces_characterization_2012,
	title        = {Characterization of mechanical properties of a hollow cylinder with zero group velocity {Lamb} modes},
	author       = {Cès, M. and Royer, D. and Prada, C.},
	year         = 2012,
	month        = jul,
	journal      = {J. Acoust. Soc. Am.},
	volume       = 132,
	number       = 1,
	pages        = {180--185},
	doi          = {10.1121/1.4726033},
	issn         = {0001-4966},
}

@article{clorennec_local_2007,
	title        = {Local and noncontact measurements of bulk acoustic wave velocities in thin isotropic plates and shells using zero group velocity {Lamb} modes},
	author       = {Clorennec, Dominique and Prada, Claire and Royer, Daniel},
	year         = 2007,
	month        = feb,
	journal      = {J. Appl. Phys.},
	volume       = 101,
	number       = 3,
	pages        = {034908},
	doi          = {10.1063/1.2434824},
	issn         = {0021-8979},
}

@article{grunsteidl_determination_2018,
	title        = {Determination of thickness and bulk sound velocities of isotropic plates using zero-group-velocity {Lamb} waves},
	author       = {Grünsteidl, Clemens and Berer, Thomas and Hettich, Mike and Veres, István},
	year         = 2018,
	month        = jun,
	journal      = {Appl. Phys. Lett.},
	volume       = 112,
	number       = 25,
	pages        = 251905,
	doi          = {10.1063/1.5034313},
	issn         = {0003-6951},
}

@article{gravenkamp_numerical_2012,
	title        = {A numerical approach for the computation of dispersion relations for plate structures using the {Scaled} {Boundary} {Finite} {Element} {Method}},
	author       = {Gravenkamp, Hauke and Song, Chongmin and Prager, Jens},
	year         = 2012,
	month        = may,
	journal      = {J. Sound Vib.},
	volume       = 331,
	number       = 11,
	pages        = {2543--2557},
	doi          = {10.1016/j.jsv.2012.01.029},
	issn         = {0022-460X},
}

@article{gravenkamp_high-order_2021,
	title        = {High-{Order} {Shape} {Functions} in the {Scaled} {Boundary} {Finite} {Element} {Method} {Revisited}},
	author       = {Gravenkamp, Hauke and Saputra, Albert A. and Duczek, Sascha},
	year         = 2021,
	month        = mar,
	journal      = {Arch. Comput. Methods Eng.},
	volume       = 28,
	number       = 2,
	pages        = {473--494},
	doi          = {10.1007/s11831-019-09385-1},
	issn         = {1134-3060},
}

@software{kiefer_gew_2022,
	title        = {{GEW} dispersion script},
	author       = {Kiefer, Daniel A.},
	year         = 2022,
	month        = aug,
	doi          = {10.5281/zenodo.7010603},
	url          = {https://github.com/dakiefer/GEW_dispersion_script},
	copyright    = {ISC License, Open Access},
	note         = {{DOI}: \href{https://dx.doi.org/10.5281/zenodo.7010603}{10.5281/zenodo.7010603}}
}

@software{plestenjak_gew_2023,
	title        = {{GEW} {ZGV} computation},
	author       = {Plestenjak, Bor and Kiefer, Daniel A.},
	year         = 2023,
	month        = jan,
	doi          = {10.5281/zenodo.7537441},
	url          = {https://github.com/dakiefer/GEW_ZGV_computation},
	copyright    = {ISC License, Open Access},
	note         = {{DOI}: \href{https://dx.doi.org/10.5281/zenodo.7537441}{10.5281/zenodo.7537441}}
}

@article{kiefer_computing_2023,
	title        = {Computing zero-group-velocity points in anisotropic elastic waveguides: {Globally} and locally convergent methods},
	shorttitle   = {Computing zero-group-velocity points in anisotropic elastic waveguides},
	author       = {Kiefer, Daniel A. and Plestenjak, Bor and Gravenkamp, Hauke and Prada, Claire},
	year         = 2023,
	month        = feb,
	journal      = {J. Acoust. Soc. Am.},
	volume       = 153,
	number       = 2,
	pages        = {1386--1398},
	doi          = {10.1121/10.0017252},
	issn         = {0001-4966}
}

@article{hopcroft_what_2010,
	title        = {What is the {Young}'s {Modulus} of {Silicon}?},
	author       = {Hopcroft, Matthew A. and Nix, William D. and Kenny, Thomas W.},
	year         = 2010,
	month        = apr,
	journal      = {J. Microelectromech. Syst.},
	volume       = 19,
	number       = 2,
	pages        = {229--238},
	doi          = {10.1109/JMEMS.2009.2039697},
	issn         = {1057-7157, 1941-0158}
}

@article{prada_influence_2009,
	title        = {Influence of the anisotropy on zero-group velocity {Lamb} modes},
	author       = {Prada, Claire and Clorennec, Dominique and Murray, Todd W. and Royer, Daniel},
	year         = 2009,
	month        = aug,
	journal      = {J. Acoust. Soc. Am.},
	volume       = 126,
	number       = 2,
	pages        = {620--625},
	doi          = {10.1121/1.3167277},
	issn         = {0001-4966},
}

@article{masserey_defect_2019,
	title        = {Defect detection in monocrystalline silicon wafers using high frequency guided waves},
	author       = {Masserey, Bernard and Simon, Mathieu and Robyr, Jean-Luc and Fromme, Paul},
	year         = 2019,
	month        = may,
	journal      = {AIP Conference Proceedings},
	volume       = 2102,
	number       = 1,
	pages        = {020013},
	doi          = {10.1063/1.5099717},
	issn         = {0094-243X}
}

@article{maznev_mapping_2011,
	title        = {Mapping the band structure of a surface phononic crystal},
	author       = {Maznev, A A and Wright, O B and Matsuda, O},
	year         = 2011,
	month        = jan,
	journal      = {New J. Phys.},
	volume       = 13,
	number       = 1,
	pages        = {013037},
	doi          = {10.1088/1367-2630/13/1/013037},
	issn         = {1367-2630}
}

@article{chakrapani_crack_2012,
	title        = {Crack {Detection} in {Full} {Size} {Cz}-{Silicon} {Wafers} {Using} {Lamb} {Wave} {Air} {Coupled} {Ultrasonic} {Testing} ({LAC}-{UT})},
	author       = {Chakrapani, Sunil Kishore and Padiyar, M. Janardhan and Balasubramaniam, Krishnan},
	year         = 2012,
	month        = mar,
	journal      = {J. Nondestr. Eval.},
	volume       = 31,
	number       = 1,
	pages        = {46--55},
	doi          = {10.1007/s10921-011-0119-3},
	issn         = {1573-4862}
}

@article{belyaev_crack_2006,
	title        = {Crack detection and analyses using resonance ultrasonic vibrations in full-size crystalline silicon wafers},
	author       = {Belyaev, A. and Polupan, O. and Dallas, W. and Ostapenko, S. and Hess, D. and Wohlgemuth, J.},
	year         = 2006,
	month        = mar,
	journal      = {Appl. Phys. Lett.},
	volume       = 88,
	number       = 11,
	pages        = 111907,
	doi          = {10.1063/1.2186393},
	issn         = {0003-6951}
}

@article{song_crack_2013,
	title        = {Crack {Detection} in {Single}-{Crystalline} {Silicon} {Wafer} {Using} {Laser} {Generated} {Lamb} {Wave}},
	author       = {Song, Min-Kyoo and Jhang, Kyung-Young},
	year         = 2013,
	month        = dec,
	journal      = {Adv. Mater. Sci. Eng.},
	volume       = 2013,
	pages        = {e950791},
	doi          = {10.1155/2013/950791},
	issn         = {1687-8434},
}

@article{ryzy_determining_2023,
	title        = {Determining longitudinal and transverse elastic wave attenuation from zero-group-velocity {Lamb} waves in a pair of plates},
	author       = {Ryzy, Martin and Veres, István and Berer, Thomas and Salfinger, Michael and Kreuzer, Susanne and Yan, Guqi and Scherleitner, Edgar and Grünsteidl, Clemens},
	year         = 2023,
	month        = apr,
	journal      = {J. Acoust. Soc. Am.},
	volume       = 153,
	number       = 4,
	pages        = 2090,
	doi          = {10.1121/10.0017652},
	issn         = {0001-4966}
}

@article{chapuis_excitation_2010,
	title        = {Excitation and focusing of {Lamb} waves in a multilayered anisotropic plate},
	author       = {Chapuis, Bastien and Terrien, Nicolas and Royer, Daniel},
	year         = 2010,
	month        = jan,
	journal      = {J. Acoust. Soc. Am.},
	volume       = 127,
	number       = 1,
	pages        = {198--203},
	doi          = {10.1121/1.3263607},
	issn         = {0001-4966}
}

@article{karmazin_study_2013,
	title        = {A study of time harmonic guided {Lamb} waves and their caustics in composite plates},
	author       = {Karmazin, Alexander and Kirillova, Evgenia and Seemann, Wolfgang and Syromyatnikov, Pavel},
	year         = 2013,
	month        = jan,
	journal      = {Ultrasonics},
	volume       = 53,
	number       = 1,
	pages        = {283--293},
	doi          = {10.1016/j.ultras.2012.06.012},
	issn         = {0041-624X},
}

@article{biot_general_1957,
	title        = {General {Theorems} on the {Equivalence} of {Group} {Velocity} and {Energy} {Transport}},
	author       = {Biot, M. A.},
	year         = 1957,
	month        = feb,
	journal      = {Phys. Rev.},
	volume       = 105,
	number       = 4,
	pages        = {1129--1137},
	doi          = {10.1103/PhysRev.105.1129}
}

@article{yantchev_thin-film_2011,
	title        = {Thin-film zero-group-velocity {Lamb} wave resonator},
	author       = {Yantchev, Ventsislav and Arapan, Lilia and Katardjiev, Ilia and Plessky, Victor},
	year         = 2011,
	month        = jul,
	journal      = {Appl. Phys. Lett.},
	volume       = 99,
	number       = 3,
	pages        = {033505},
	doi          = {10.1063/1.3614559},
	issn         = {0003-6951}
}

@article{caliendo_zero-group-velocity_2017,
	title        = {Zero-group-velocity acoustic waveguides for high-frequency resonators},
	author       = {Caliendo, C. and Hamidullah, M.},
	year         = 2017,
	month        = nov,
	journal      = {J. Phys. D: Appl. Phys.},
	volume       = 50,
	number       = 47,
	pages        = 474002,
	doi          = {10.1088/1361-6463/aa900f},
	issn         = {0022-3727},
}

@article{hackett_aluminum_2023,
	title        = {Aluminum scandium nitride films for piezoelectric transduction into silicon at gigahertz frequencies},
	author       = {Hackett, L. and Miller, M. and Beaucejour, R. and Nordquist, C. M. and Taylor, J. C. and Santillan, S. and Olsson, R. H. and Eichenfield, M.},
	year         = 2023,
	month        = aug,
	journal      = {Appl. Phys. Lett.},
	volume       = 123,
	number       = 7,
	pages        = {073502},
	doi          = {10.1063/5.0151434},
	issn         = {0003-6951}
}

@article{bramhavar_negative_2011,
	title        = {Negative refraction and focusing of elastic {Lamb} waves at an interface},
	author       = {Bramhavar, Suraj and Prada, Claire and Maznev, Alexei A. and Every, Arthur G. and Norris, Theodore B. and Murray, Todd W.},
	year         = 2011,
	month        = jan,
	journal      = {Phys. Rev. B},
	volume       = 83,
	number       = 1,
	pages        = {014106},
	doi          = {10.1103/PhysRevB.83.014106}
}

@article{legrand_cloaking_2021,
	title        = {Cloaking, trapping and superlensing of lamb waves with negative refraction},
	author       = {Legrand, François and Gérardin, Benoît and Bruno, François and Laurent, Jérôme and Lemoult, Fabrice and Prada, Claire and Aubry, Alexandre},
	year         = 2021,
	month        = dec,
	journal      = {Sci. Rep.},
	volume       = 11,
	number       = 1,
	pages        = 23901,
	doi          = {10.1038/s41598-021-03146-6},
	issn         = {2045-2322}
}

@book{rupitsch_piezoelectric_2018,
	title        = {Piezoelectric {Sensors} and {Actuators} - {Fundamentals} and {Applications}},
	shorttitle   = {Piezoelectric {Sensors} and {Actuators}},
	author       = {Rupitsch, Stefan J.},
	year         = 2018,
	month        = aug,
	publisher    = {Springer},
	address      = {Berlin},
	series       = {Topics in {Mining}, {Metallurgy} and {Materials} {Engineering}},
	isbn         = {978-3-662-57532-1},
	edition      = 1
}

\section*{Acknowledgements}
\textbf{Funding:} This work has received support under the program «Investissements d’Avenir» launched by the French Government under Reference No. ANR-10-LABX-24.

% \section*{Author contributions}% bigskip need due to bug in template
\noindent
\textbf{Author contributions:} DAK and CP conceived and designed the study. SM conceived and performed the experiment and the data curation. DAK performed the modeling and implemented the numerical model. All authors analyzed and interpreted the experimental results. DAK and SM wrote the manuscript. All authors proofread and revised the manuscript.

% \section*{Competing interests}
\noindent
\textbf{Competing interests:} The authors declare no competing interests.

\noindent
\textbf{Data and material availability:} All data needed to evaluate the conclusions in the paper are present in the paper and/or the Supplementary Materials. Our code is provided at \url{https://dx.doi.org/10.5281/zenodo.10114244}.

\section*{Supplementary Materials}
We provide four supplementary videos showing the spatio-temporal dynamics of the studied waves.

\end{document}